\documentclass[a4paper,onecolumn,10pt]{edpsci}
\usepackage{amsmath}
\usepackage{graphicx}
\usepackage{epstopdf}
\usepackage{hyperref}
\usepackage{url}
\hypersetup{colorlinks=true,citecolor=blue,urlcolor=blue,linkcolor=blue}

\journalname{Acta Acustica}

\articletype{Scientific article}

\usepackage[T1]{fontenc} 
\usepackage[american]{babel}

 \usepackage{caption}
 \usepackage{subcaption}

\usepackage{tabularx}
\usepackage{booktabs}

\usepackage{setspace}

\usepackage[sorting=none, giveninits=true, maxbibnames=99]{biblatex}
\bibliography{manual_bib.bib}
\AtEveryBibitem{
\ifentrytype{article}{%
    \clearfield{url}%
    \clearfield{doi}%
    }{}%
\ifentrytype{inproceedings}{%
    \clearfield{url}%
    \clearfield{doi}%
    }{}%
\ifentrytype{book}{%
    \clearfield{url}%
    \clearfield{doi}%
    }{}%
\ifentrytype{incollection}{%
    \clearfield{url}%
    \clearfield{doi}%
    }{}%
    \clearlist{language} 
    \clearfield{urlyear} 
    \clearfield{note}%
    \clearfield{issn}%
    \clearfield{isbn}%
    \clearfield{eprint}%
}  
\usepackage{csquotes}

\title{Subjective quality evaluation of personalized own voice\\reconstruction systems}

\institute{Fraunhofer Institute for Digital Media Technology IDMT,
Oldenburg Branch for Hearing, Speech and Audio Technology HSA, Germany \and
 Carl von Ossietzky Universität Oldenburg, Department of Medical Physics and Acoustics and Cluster of Excellence Hearing4all, Germany}

\author{Mattes Ohlenbusch\inst{1}\correspondingauthor{\email{mattes.ohlenbusch@idmt.fraunhofer.de}} \and Christian Rollwage\inst{1} \and Simon Doclo\inst{1,2} \and Jan Rennies\inst{1,2}}

\begin{document}

\abstract{
\doublespacing
\noindent Own voice pickup technology for hearable devices facilitates communication in noisy environments.
Own voice reconstruction (OVR) systems enhance the quality and intelligibility of the recorded noisy own voice signals. 
Since disturbances affecting the recorded own voice signals depend on individual factors, personalized OVR systems have the potential to outperform generic OVR systems.
In this paper, we propose personalizing OVR systems through data augmentation and fine-tuning, comparing them to their generic counterparts. 
We investigate the influence of personalization on speech quality assessed by objective metrics and conduct a subjective listening test to evaluate quality under various conditions.
In addition, we assess the prediction accuracy of the objective metrics by comparing predicted quality with subjectively measured quality.
Our findings suggest that personalized OVR provides benefits over generic OVR for some talkers only.
Our results also indicate that performance comparisons between systems are not always accurately predicted by objective metrics. 
In particular, certain disturbances lead to a consistent overestimation of quality compared to actual subjective ratings. 
} 


\maketitle

\doublespacing


\section{Introduction}
\label{sec:introduction}
Speech communication is often impaired in noisy environments. 
In-the-ear hearable devices, i.e., smart earpieces with a loudspeaker and one or more microphones, can be used to improve communication in such environments, e.g., by capturing and transmitting the user's own voice to a mobile phone or another hearable~\cite{bouserhal2013integration, nordholm_assistive_2015}.
Here, we consider the scenario where a hearable with an outer and an in-ear microphone aims to capture the user’s own voice, e.g., to be transmitted via a wireless link to another hearable or a mobile phone. 
The outer microphone captures environmental noise along with recording the own voice. 
While the in-ear microphone benefits from the attenuation of environmental noise due to ear canal occlusion, the recorded own voice suffers from low-frequency amplification (below ca.~1\,kHz), band-limitation (above ca.~2\,kHz), and body-produced noise~\cite{bouserhal_-ear_2019, gaardbaek_origin_2025}.
Own voice recorded at the in-ear microphone consists of an air-conducted and a body-conducted component.
The air-conducted component strongly depends on the tightness of the fit in the ear canal.
The amount of body-conducted own voice recorded at the in-ear microphone depends on hearable device properties, such as device fit and insertion depth~\cite{hansen_occlusion_1998-1, zurbrugg_investigations_2014},
individual anatomic factors such as residual ear canal volume and shape~\cite{stenfelt_model_2007, vogl_individualized_2019}, the generated sounds or phonemes being uttered~\cite{reinfeldt_hearing_2010, saint-gaudens_towards_2022}, and mouth movements~\cite{richard_change_2025}.
Environmental noise recorded at the in-ear microphone also varies with the device fit to the individual ear shape~\cite{denk_hearpiece_2021}. 

For communication applications, an own voice reconstruction (OVR) system is needed in order to reconstruct own voice from noisy hearable signals. 
Previous traditional signal processing OVR approaches are based on e.g., equalization filter design~\cite{kondo_equalization_2006}, statistical modeling~\cite{shin_priori_2015}, or non-linear bandwidth extension of in-ear own voice signals~\cite{bouserhal_-ear_2017}.
More recently proposed deep neural network (DNN)-based OVR systems are commonly designed to work for multiple potential device users~\cite{wang_fusing_2022, hauret_configurable_2023, ohlenbusch_multi-microphone_2024, li_two-stage_2024, li_restoration_2024}\footnote{
Although some of these approaches have been proposed and validated for body-conduction microphones, they can also be applied to in-ear microphones.}, 
which is achieved by training them with data from multiple talkers in order to achieve robustness to individual variation.
Since such systems are not specific to any particular user, in this work we refer to them as generic systems. 
However, since the degradations affecting noisy hearable signals are subject to several individual factors, personalized OVR systems could provide a benefit over generic systems by accounting for individual differences.

In~\cite{edraki_speaker_2024}, it has been proposed to personalize an OVR system by first training a generic system, and then fine-tuning the system incorporating speaker identification information into the system.
By comparison, the personalized systems achieved higher reconstruction performance than generic systems in metrics predicting quality and intelligibility.
Similarly, in~\cite{sui_tramba_2024} it has been proposed to personalize an OVR system for smart glasses by first pre-training a generic system for bandwidth extension of band-limited speech, and then fine-tuning the system using few recorded body-conduction signals. 
The personalized system achieved higher reconstruction performance in predictive metrics compared to the pre-trained generic system and compared to generic systems trained with own voice signals of multiple talkers (without pre-training).
While in both~\cite{edraki_speaker_2024} and~\cite{sui_tramba_2024} personalization yields a benefit over generic systems, it is not yet known whether personalized training before fine-tuning can yield additional benefits in performance over pre-training a generic system. In addition, quality predictions by objective metrics often do not correlate well with subjective ratings of body-conducted speech~\cite{richard_comparison_2023}. 
To our knowledge, a formal subjective evaluation study of personalized OVR systems has not yet been presented.

%
The present study therefore extends previous work: 
We train generic and personalized OVR systems and compare them using objective metrics and systematic subjective ratings.
We investigate both generic and personalized data augmentation for pre-training, and both generic and personalized fine-tuning.
With respect to evaluation methodology, this study aims to provide insights into which instrumental metrics are most suitable for assessing OVR performance.
OVR may be a particularly challenging case for performance assessment, since depending on the approach and microphone signals used, the metrics may need to be able to account for the effects of bandwidth limitation and extension.

\section{Own voice reconstruction} 
\label{sec:ovr}

\subsection{Signal model}
\label{sec:sigmodel}
\begin{figure}
    \centering
\includegraphics{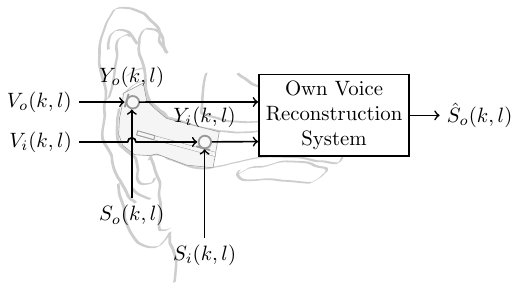} 
    \caption{Block diagram of own voice reconstruction using an outer and an in-ear microphone of a hearable.}
    \label{fig:signal_model_diagram}
\end{figure}
We consider a hearable device equipped with an outer microphone and an in-ear microphone, as depicted in Fig.~\ref{fig:signal_model_diagram}.
The microphone signals are denoted by subscripts $o$ for the outer microphone and $i$ for the in-ear microphone. 
We assume that the hearable is worn by a talker in a noisy environment.
In the short-time Fourier transform (STFT) domain, $S_o(k,l)$ and $S_i(k,l)$ denote the own voice signals of the talker at both microphones, where $k$ and $l$ denote the frequency index and the frame index.
The noisy outer and in-ear microphone signals are given by 
\begin{align}
    Y_o(k,l) & = S_o(k,l) + V_o(k,l)\;,\label{eq:sigmodel_outer}\\
    Y_i(k,l) & = S_i(k,l) + V_i(k,l)\;,\label{eq:sigmodel_inear}
\end{align}
where the noise components are denoted by $V_o(k,l)$ and $V_i(k,l)$. 
We assume the noise components predominantly consist of environmental noise at both microphones, but also microphone self-noise with a much lower level at both microphones and additional body-produced noise at the in-ear microphone.

\subsection{Generic and personalized own voice reconstruction systems} 
\label{sec:gen_and_pers_ovr_systems}
The goal of own voice reconstruction is to estimate the clean own voice signal $S_o(k,l)$ from the noisy outer and in-ear microphone signals using a DNN-based OVR system $\mathcal{D}$, i.e., 
\begin{equation}
    \hat{S}_o(k,l) = \mathcal{D}\{Y_o(k,l), Y_i(k,l)\}\;.
\end{equation}
If the OVR system is trained to be able to reconstruct own voice of multiple talkers, we refer to the system as generic.
For noisy own voice signals $Y_o^a(k,l)$ and $Y_i^a(k,l)$, the same generic system $\bar{\mathcal{D}}$ is used for any target talker $a$, i.e.,
\begin{equation}
\label{eq:gen_ovr}
    \hat{S}_o(k,l) = \bar{\mathcal{D}}\{Y_o^a(k,l), Y_i^a(k,l)\}\;.
\end{equation}
In this case, neither the system $\bar{\mathcal{D}}$ nor its output $\hat{S}_o$ are personalized for talker $a$.
Training a generic OVR system requires a sufficient amount of training data from multiple target talkers. 
The distortions affecting the in-ear own voice signal (e.g., band-limitation, low-frequency attenuation) depend on individual factors such as fit quality and insertion depth.
As a result, a generic OVR system may not achieve the same quality for all talkers.

In this work, we propose to train an OVR system to reconstruct own voice of a single target talker.
For a target talker $a$, the corresponding personalized OVR system $\mathcal{D}^a$ aims to reconstruct the speech of this particular talker, i.e.,
\begin{equation}
\label{eq:pers_ovr}
    \hat{S}_o^a(k,l) = \mathcal{D}^a\{Y_o^a(k,l), Y_i^a(k,l)\}\;.
\end{equation}
The personalized system produces a personalized output $\hat{S}_o^a$.
However, obtaining such a personalized OVR system requires a sufficient amount of talker-specific training data.

To investigate personalized own voice reconstruction, we compare generic and personalized variants of the same DNN architecture originally proposed in~\cite{tesch_insights_2023}. 
The architecture is referred to as frequency- and time-domain joint nonlinear filter (FT-JNF).
This architecture has previously been applied to OVR in~\cite{ohlenbusch_multi-microphone_2024, ohlenbusch_speech-dependent_2024, ohlenbusch_lowcomplexity_2025}.
In this work, we use a variant with 256 hidden units in the first and 128 hidden units in the second LSTM layer, respectively, leading to a DNN size of 466\,k parameters with a complexity of 7.55\,G\,MACs (Multiply-Accumulate operations) per second.

\subsection{Training-based personalization}
\label{sec:trainingbased_pers}
In order to train an OVR system using both an outer and an in-ear microphone, multi-channel own voice signals are required as training data.
A data augmentation method based on phoneme-specific relative transfer functions (RTFs) was proposed in~\cite{ohlenbusch_speech-dependent_2024}.
The method was shown to improve OVR performance compared to only using a small recorded dataset of own voice signals directly for training.

\subsubsection{Generic own voice data augmentation} 
\label{sec:gen_ov_da}
In~\cite{ohlenbusch_modeling_2023}, a method was proposed to simulate in-ear own voice signals from outer microphone signals, enabling generation of additional training utterances not included in the recorded dataset.
From the recorded dataset, phoneme-specific RTFs between the outer and the in-ear microphone are estimated for each recorded talker $a$, over all time frames in which a specific phoneme $p$ occurs.
The estimated individual, phoneme-specific RTFs are denoted by $\hat{H}_{p}^a(k)$. 
We refer to a set of all phoneme-specific RTFs from a single talker $a$ as the transfer characteristics model of talker $a$.
For RTF estimation, we assume the small recorded dataset contains no environmental noise. 
Sensor noise and body-produced noise are assumed negligible compared to the own voice.

For simulation, an outer microphone signal $S_o(k,l)$ of a random different talker
is phoneme-annotated to obtain the phoneme annotation sequence $p_o(l)$, where $p_o(l$ denotes the phoneme at frame $l$.
The simulated in-ear signal $\hat{S}_i^a(k,l)$ of a random talker is then obtained as
\begin{equation}
    \hat{S}_i(k,l) =  \hat{H}_{p_{o}(l)}(k) \times S_o(k,l)\;,
    \label{eq:augmentation_speechdep}
\end{equation} 
where $\times$ denotes multiplication.
Additionally, to avoid artifacts during phoneme transitions, temporal smoothing of $\hat{H}_{p_{o}(l)}(k)$ is carried out (see~\cite{ohlenbusch_modeling_2023} for details).   
Instead of assuming recorded outer microphone signals are available, it was proposed in~\cite{ohlenbusch_speech-dependent_2024} to use clean speech signals from standard datasets instead.
An OVR system can then be trained with augmented own voice signals, consisting of clean speech signals used as the outer microphone own voice signal and the corresponding simulated in-ear own voice signal.
Since standard datasets are readily available, this method allows for the simulation of a large amount of simulated in-ear own voice signals.
In~\cite{ohlenbusch_speech-dependent_2024}, a transfer characteristic model from a random talker was selected for each new utterance to obtain a simulated dataset of multiple talkers that can be used to train a generic OVR system.

\subsubsection{Personalized own voice data augmentation}
\label{sec:pers_ov_da}
A personalized OVR system can be trained based on personalized own voice data augmentation. For this purpose, we propose to perform augmentation using only a single transfer characteristic model of a single target talker $a$. 
Similar to~\eqref{eq:augmentation_speechdep}, the simulated in-ear signal $\hat{S}_i^a(k,l)$ of target talker $a$ is obtained as
\begin{equation}
    \hat{S}_i^a(k,l) =  \hat{H}_{p_{o}(l)}^a(k) \times S_o(k,l)\;. 
    \label{eq:augmentation_speechdep_perso}
\end{equation} 
This way, an entire personalized own voice dataset for talker $a$ can be simulated.

\subsubsection{Fine-tuning}
\label{sec:fine-tuning}
After training an OVR system with augmented own voice signals, the recorded own voice signals can be used to fine-tune the system, further improving performance.
In~\cite{ohlenbusch_speech-dependent_2024}, it was shown that fine-tuning of the entire OVR system is more beneficial compared to only fine-tuning parts of the system.
Fine-tuning is carried out with a smaller initial learning rate than the learning rate used for training with augmented signals.
In order to fine-tune a generic system, it is possible to apply either generic or personalized fine-tuning.
For \textit{generic fine-tuning}, the recorded own voice signals of all available talkers are used.
Since this procedure aims to reconstruct the own voice of multiple individual talkers, the fine-tuned OVR system is generic.
For \textit{personalized fine-tuning}, the recorded own voice signals of only the target talker are used.
Since this procedure aims to reconstruct the own voice of a single individual talker, the fine-tuned OVR system is personalized.

\section{OVR system training setup} 
\label{sec:ovr_training_setup}
An almost identical training setup as described in~\cite{ohlenbusch_speech-dependent_2024} was used. 
The split of the dataset of recorded own voice signals is different (see Sect.~\ref{sec:datasets}), and in addition to generic data augmentation and fine-tuning, personalized data augmentation and fine-tuning are also considered.

\subsection{Datasets} 
\label{sec:datasets}
The evaluation uses clean recorded own voice signals made with the Hearpiece hearable prototype~\cite{denk_one-size-fits-all_2019}.
A dataset of German own voice signals of 18 talkers with 306 utterances each is split into 206 utterances for training, 50 utterances for validation, and 50 utterances for testing, respectively.
For personalized training, all training utterances of the target talker are used in data augmentation or fine-tuning.
For generic training, a random subset consisting of 206 utterances is selected from the training utterances of all 18 talkers, so that the number of recorded own voice signals is the same for generic and personalized training. 
The augmented own voice signals are obtained by augmenting 10\% of the German portion of the CommonVoice dataset~\cite{ardila_common_2020} (v11.0), which corresponds to 115.7 hours of speech signals.

%
The noise signals at both microphones used for training and evaluating the OVR systems based on instrumental metrics
are a spatialized version of the fifth DNS challenge~\cite{dubey_icassp_2024}, obtained following the procedure in~\cite{ohlenbusch_multi-microphone_2024} using individually matched, measured transfer functions for the same users as in the dataset of recorded own voice signals. 
Measurements from 8 horizontal directions in 45°-steps at a distance of 1.5\,m are used to compute either point source signals (single direction) or pseudo-diffuse noise signals (8 directions). 

\subsection{Evaluation details}
\label{sec:eval_details}
The evaluation was carried out similarly to~\cite{ohlenbusch_speech-dependent_2024}. For evaluation, OVR performance was evaluated at fixed signal-to-noise ratios (SNRs) of -10, -5, 0, 5, and 10\,dB, and results were averaged over test set examples and the considered SNRs. 
The performance was assessed in terms of instrumental metrics (see Sect.~\ref{sec:objective_metrics}).
In order to investigate the influence of personalized OVR, the performance of generic and personalized systems was compared for each talker in the recorded dataset. 
For generic systems, a single system was trained with noisy own voice signals from all 18 talkers. The generic system was then evaluated separately on each talker's test set (excluding training and validation utterances).
For personalized systems, a separate personalized system was trained for each talker. 
Each personalized system was then evaluated only on the test set of the same corresponding talker, respectively.

\section{Objective quality prediction metrics}
\label{sec:objective_metrics}
Since it is currently unknown which metrics are best suited for predicting OVR performance, a variety of instrumental metrics are tested against subjective ratings.
OVR systems are commonly evaluated using speech enhancement metrics, which aim to predict speech quality, intelligibility, or listening effort.
Some of the metrics require a reference signal (intrusive metrics), which is a clean speech signal corresponding to the speech content in the noisy or processed signal to be evaluated by the metric.
In this work, the following intrusive metrics are investigated:
\begin{description}
    \item[PESQ] Wideband perceptual evaluation of speech quality (PESQ) \cite{international_telecommunications_union_itu_itu-t_2001} is a metric predicting speech quality.
    Although it is often used beyond its original scope and has since been superseded by POLQA~\cite{international_telecommunications_union_itu_itu-t_2018}, it remains a popular metric to evaluate speech enhancement systems. 
    In particular, in~\cite{santos2016objective} it was observed that PESQ predictions correlate well with subjective ratings of speech recorded at an in-ear microphone, and slightly better than POLQA.
    When predicting quality of bandwidth-extended signals in~\cite{Bauer2015artificial} PESQ showed low correlations, although slightly better than POLQA.
    The scale of PESQ output values ranges from 0.5 to 4.5.
    \item[ESTOI] Extended short-time objective intelligibility (ESTOI) was originally designed to predict intelligibility. Strictly, this requires a score transformation based on the evaluation material~\cite{jensen_algorithm_2016}. 
    Nevertheless, it is often used without transformation to evaluate the performance of speech enhancement systems. 
    In~\cite{hauret_configurable_2023}, the original STOI~\cite{taal_algorithm_2011} was observed to predict subjective quality ratings well. 
    The extended STOI is designed to work for a wider scope than the original STOI, including highly modulated noise. 
    The scale of raw ESTOI output values ranges from 0 to 1.  
    \item[eMoBi-Q] The efficient model for binaural audio quality (eMoBi-Q) combines the monaural generalized power spectrum model for quality GPSM$^\mathrm{q}$~\cite{biberger2018gpsmq}
    and a binaural model and simplifies some computational aspects of them while maintaining accuracy for predicting quality in hearing device applications~\cite{eurich_computationally_2024}.
    For monaural signals, eMoBi-Q represents a computationally efficient version of GPSM$^\mathrm{q}$.
    The scale of predicted values by eMoBi-Q ranges from 0 to 1.
    \item [PEMO-Q~PSM] An audio quality model which is based on a psychoacoustically validated auditory perception model (PEMO)~\cite{dau1996quantitative} is PEMO-Q~\cite{huber_pemoq_2006}. 
    In this study, its most basic metric, i.e., the Perceptual Similarity Measure (PSM), is used. 
    PSM is the linear cross correlation between the internal representations of a pair of test and reference signals. The internal representations are computed by the perception model. 
    In this case, the simpler variant is used in which internal representations are obtained by modeling amplitude modulation processing with a low-pass filter. 
    A voice activity detector is used to select signal segments containing speech before processing the signal by the model. 
    Being a Pearson correlation coefficient, PSM can assume values between -1 and 1, with 1 indicating perceptual identity between test and reference signal (interpreted as maximum quality of the test signal); practically, output values are in the range between 0 and 1. 
    \item[SCOREQ distance] Speech contrastive regression for quality (SCOREQ)~\cite{ragano2024scoreq} is a method for predicting speech quality designed to capture the continuous nature of the MOS scale.
    When applied to signals for which a reference signal is available, a distance value can be computed indicating similarity to the reference signal. 
    The distance ranges from 0 to infinity, with smaller values indicating higher similarity.
\end{description} 
In this work, the reference signal for all intrusive metrics is the clean outer microphone signal.
A higher value indicates better quality for all metrics except SCOREQ distance, where a lower distance value indicates better quality. 
In contrast to intrusive metrics, non-intrusive metrics do not require a reference signal. Many popular non-intrusive metrics are based on deep learning and are trained to predict subjective ratings from noisy or processed signals only. 
In this work, the following non-intrusive metrics are investigated:
\begin{description}
    \item[DNSMOS] Deep Noise Suppression Mean Opinion Score (DNSMOS)~\cite{reddy_dnsmos_2022} is a non-intrusive DNN-based metric to predict quality of an audio signal in terms of speech/signal quality (DNSMOS~SIG), background quality (DNSMOS~BAK), and overall quality (DNSMOS~OVRL) based on P.835~\cite{itu-t_2018_p835}, as well as overall quality based on P.808~\cite{itu-t_2018_p808}, here referred to as DNSMOS~P808. 
    In both cases, DNSMOS predicts values on the MOS scale from 1 to 5.
    \item[SCOREQ MOS] different from the SCOREQ distance, the SCOREQ model can also be used without a reference signal to predict values on the MOS scale from 1 to 5~\cite{ragano2024scoreq}.
    \item[WV-MOS] 
    Band-limited and bandwidth-extended signals are hard to evaluate in terms of quality using objective metrics designed for full-band signals~\cite{richard_comparison_2023, Bauer2015artificial}.
    To address this, it was proposed in~\cite{andreev_hifipp_2023} to predict MOS ratings using a DNN-based approach based on a pretrained wav2vec2.0 (WV-MOS). 
    WV-MOS predicts values on the MOS scale from 1 to 5.
    \item[LEAP] The model for Listening Effort prediction from Acoustic Parameters (LEAP)~\cite{huber_leap_2018, huber_leap_itg2018} is based on an automatic phoneme recognizer for German speech. 
    Speech degradations, such as additive noise, distortions, or reverberation, increase the uncertainty of the recognition process, which has been found to correlate very well with human ratings of perceived listening effort. 
    LEAP predicts the perceived listening effort on a subjective scale, ranging from 1 (“no effort”) to 13 (“extreme effort”) and 14 (“only noise” or “no speech perceivable”)~\cite{krueger2017development}. 
\end{description}

\section{Results of instrumental assessment}
\label{sec:pesq_stoi_results}
\begin{figure*}
    \centering
    \begin{subfigure}[b]{0.49\textwidth}
        \includegraphics[width=\linewidth]{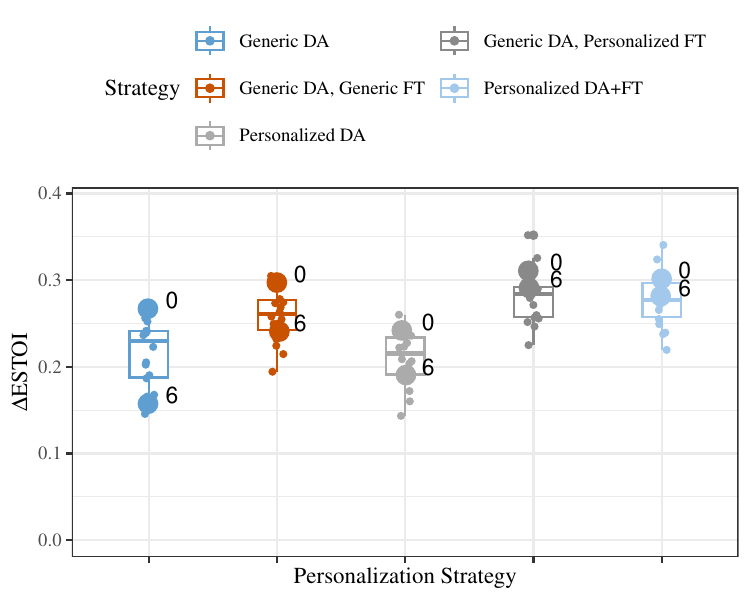}
        \caption{ESTOI improvement.} 
        \label{fig:estoi_results_boxplot}
    \end{subfigure}
    \hfill
    \begin{subfigure}[b]{0.49\textwidth}
        \includegraphics[width=\linewidth]{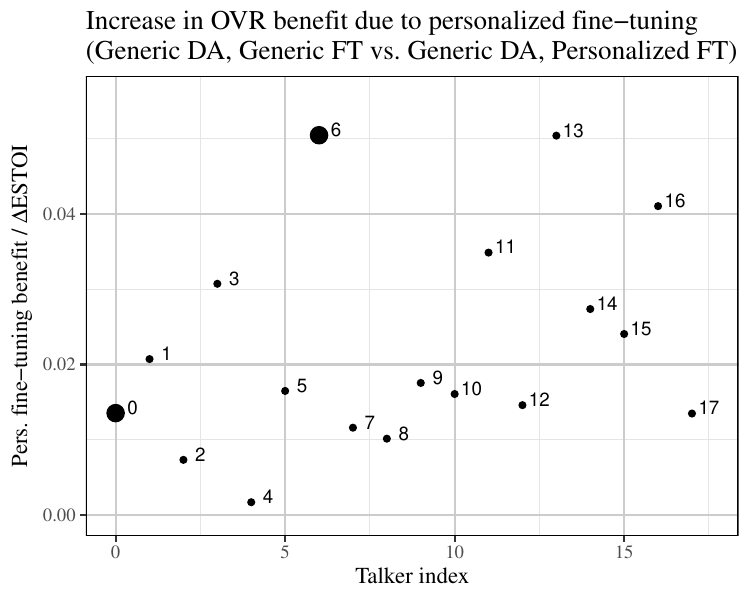}
        \caption{Predicted benefit from personalized fine-tuning.}
        \label{fig:estoi_delta_personalization}
    \end{subfigure}
    \caption{(a) ESTOI improvement achieved by OVR systems trained with different personalization strategies in data augmentation (DA) and fine-tuning (FT) and (b) individual difference in ESTOI improvement between the conditions Generic DA, generic FT and Generic DA, personalized FT. 
    Individual data points denote the average over individual target talker test sets.
    The data points labeled with 0 and 6 correspond to the talkers selected for the listening experiment as the \textit{low predicted benefit} and \textit{high predicted benefit}, respectively.}
    \label{fig:estoi_results}
\end{figure*}

\begin{figure*}
    \centering
    \begin{subfigure}[b]{0.49\textwidth}
        \includegraphics[width=\linewidth]{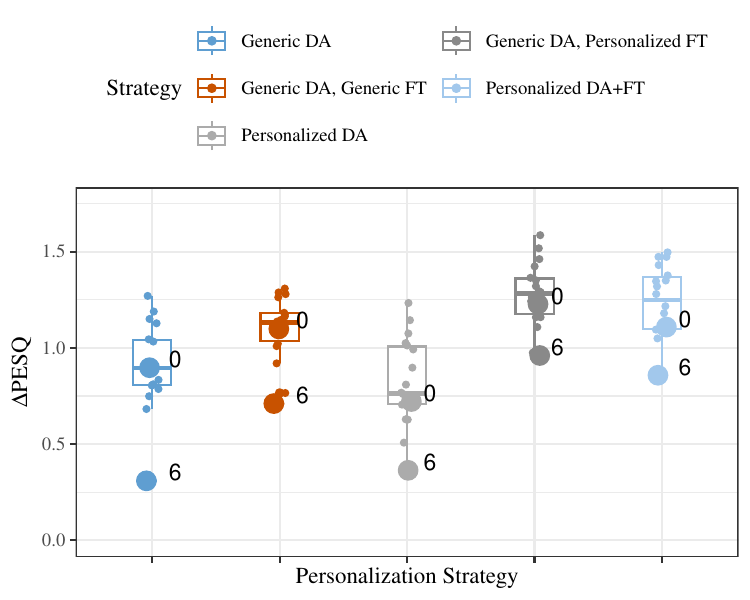}
        \caption{PESQ improvement.} 
        \label{fig:pesq_results_boxplot}
    \end{subfigure}
    \hfill
    \begin{subfigure}[b]{0.49\textwidth}
        \includegraphics[width=\linewidth]{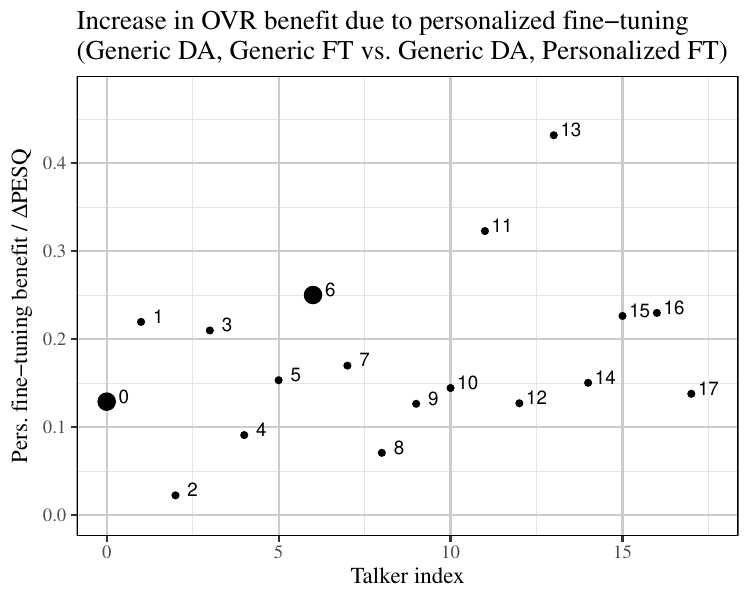}
        \caption{Predicted benefit from personalized fine-tuning.}
        \label{fig:pesq_delta_personalization}
    \end{subfigure}
    \caption{(a) PESQ improvement achieved by OVR systems trained with different personalization strategies in data augmentation (DA) and fine-tuning (FT) and (b) individual difference in PESQ improvement between the conditions generic DA, generic FT and generic DA, personalized FT. 
    Individual data points denote the average over individual target talker test sets. 
    The data points labeled with 0 and 6 correspond to the talkers selected for the listening experiment as the \textit{low predicted benefit} and \textit{high predicted benefit}, respectively.
    }
    \label{fig:pesq_results}
\end{figure*}

Figure~\ref{fig:estoi_results_boxplot} and Figure~\ref{fig:pesq_results_boxplot} show the improvements ($\Delta$) in ESTOI and PESQ, respectively, compared to the noisy outer microphone signals.

The results for individual target talker test sets are shown both as individual points as well as boxplots over all target talkers per personalization strategy.
Two exemplary target talkers (Numbers 0 and 6) are highlighted.
All strategies consistently improved both metrics across all talkers compared to the noisy outer microphone signals.
Strategies that include a fine-tuning step perform better than those without.
Personalized data augmentation without fine-tuning leads to slightly lower scores than generic fine-tuning. 
When generic data augmentation is used, personalized fine-tuning leads to better performance than generic fine-tuning.
Systems trained with generic data augmentation and then personalized fine-tuning outperformed those trained with personalized data augmentation and fine-tuning.

The results predict a consistent benefit of OVR systems personalized by fine-tuning over generic systems. 
In contrast, no consistent benefit is predicted for personalized data augmentation, which could be due to the decrease in variance in the training data relative to generic data augmentation (18 times the number of RTFs used for augmentation).  
In order to investigate whether the instrumental predictions are accurate, a subjective listening experiment is performed in the following. 

The benefit achieved by personalized over generic processing is different between individual target talkers.
In Figs.~\ref{fig:estoi_results} and~\ref{fig:pesq_results}, two example target talkers are highlighted. 
For assessing the benefit of personalized over generic fine-tuning, the difference between the $\Delta$ESTOI and $\Delta$PESQ scores in Figs.~\ref{fig:estoi_results_boxplot} and~\ref{fig:pesq_results_boxplot} for the conditions Generic DA, Generic FT an Generic DA, personalized FT are shown in Figs.~\ref{fig:estoi_delta_personalization} and~\ref{fig:pesq_delta_personalization}.
In case of talker 0, only a small additional improvement in ESTOI and PESQ is achieved by personalized fine-tuning when compared to the generic cases. 
In contrast, in case of talker 6, a large additional improvement is achieved by personalized fine-tuning in terms of ESTOI. 
While PESQ predicts a different ranking order for this talker than ESTOI, it still indicates an improvement of 0.25 $\Delta$PESQ from personalized fine-tuning compared to generic fine-tuning. 
To reduce the measurement time for the listening experiment, we chose to investigate only the performance for these two target talkers when comparing predicted quality to subjective quality ratings. 
We refer to processed signals of talker 0 as the \textit{low predicted benefit} case, and processed signals of talker 6 as the \textit{high predicted benefit} case.

\section{Listening experiment}
\label{sec:listening_experiment}

\subsection{Evaluation procedure} 
\label{sec:evaluation_procedure}
In order to investigate subjective quality achieved by OVR processing, a listening experiment based on the multiple stimulus and hidden reference (MUSHRA) standard for assessing intermediate audio quality~\cite{international_telecommunications_union_itu_itu-r_2015} was conducted. 
Unlike the MUSHRA standard, the MUSHRA-like experiment used in this paper employed different anchor signals (see Sect.~\ref{sec:processing_conditions}) than the lowpass-filtered clean speech signals defined in~\cite{international_telecommunications_union_itu_itu-r_2015}. 
The experiment was carried out using the WebMUSHRA framework~\cite{schoeffler2018webmushra}.
Participants were instructed to rate the overall quality of each signal presented to them.
Before conducting the actual experiment, a training screen was presented to the participants for familiarization.
The experiment was conducted in sound-proof listening booths, and stimuli (noisy and processed own voice) were presented over open-back headphones (Sennheiser HDA 650, calibrated for 70 dB SPL output).

\subsection{Participants} 
\label{sec:participants}
Twenty-five normal-hearing native German-speaking participants (13 female, 12 male), aged $23.8\pm3.8$\,years (mean$\pm$standard deviation), took part in the listening experiment.
All participants had pure-tone thresholds $\leq 20$\,dB hearing level at audiometric frequencies from 125\,Hz to 8\,kHz on both ears. 
One participant was excluded from the evaluation for not correctly identifying the reference signal in all the MUSHRA screens.  
All participants received hourly compensation and gave informed consent for their participation in the experiments.
The methods of the experiment were approved by the ethics committee of the
Carl von Ossietzky University of Oldenburg (protocol Drs.EK/2019/073-2). 

\subsection{Processing conditions}
\label{sec:processing_conditions} 
\begin{figure}
    \centering
    \includegraphics{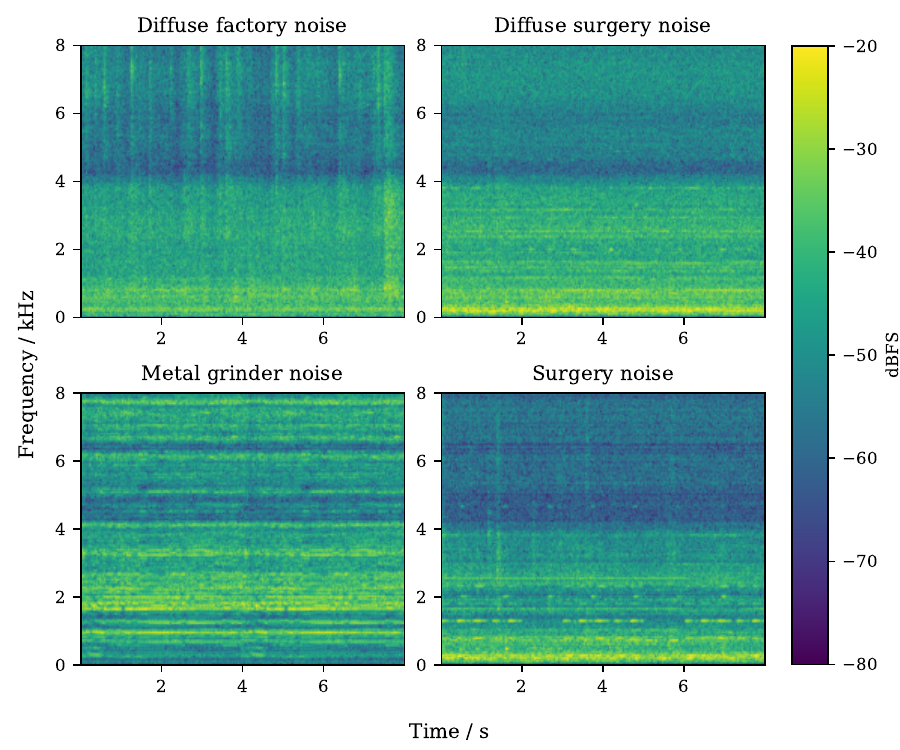} 
    \caption{Spectrograms of the recorded noise signals used in the subjective evaluation.}
    \label{fig:specgram_noise}
\end{figure}

The subjective evaluation was conducted by evaluating quality of noisy and processed own voice signals
in different processing conditions, noise types, and talkers.
The considered processing conditions were:
\begin{description}
    \item[Noisy OM] The noisy, unprocessed outer microphone signal (anchor). 
    Since the results in Sect.~\ref{sec:pesq_stoi_results} suggest a consistent improvement by OVR systems over the noisy unprocessed outer microphone signals, they were used as anchor signals for the MUSHRA-like listening test (instead of, e.g., clean band-limited signals).
    \item[Noisy IM] The noisy, unprocessed in-ear microphone signal.
    \item[MWF] An implementation of the multi-channel Wiener filter (MWF)~\cite{docloMWF} with multi-channel speech presence probability and recursive smoothing-based power spectral density estimation~\cite{souden2009gaussian-key, bagheri19_interspeech} using the noisy outer and in-ear signals as input signals and using the outer microphone as the reference microphone. MWF is used as a baseline method in the subjective evaluation.
    \item[EBEN] The extreme bandwidth extension network (EBEN)~\cite{hauret_configurable_2023} is a time-domain DNN-based system. For the listening test, it was retrained with generic data augmentation and generic fine-tuning to estimate clean outer microphone own voice signals from noisy in-ear microphone signals. 
    Training was performed using the same hyperparameters as the FT-JNF systems, including the loss function (different from the generative adversarial network training paradigm in~\cite{hauret_configurable_2023}). EBEN is used as a baseline method in the subjective evaluation.
    \item[Generic DA] The FT-JNF architecture, trained using generic data augmentation only.
    \item[Generic DA, generic FT] The FT-JNF architecture, trained using generic data augmentation and generic fine-tuning.
    \item[Generic DA, personalized FT] The FT-JNF architecture, trained using generic data augmentation and personalized fine-tuning.
    \item[Personalized DA, personalized FT] The FT-JNF architecture, trained using personalized data augmentation and personalized fine-tuning
    \item[Reference] The clean own voice signal at the outer microphone (hidden reference).
\end{description}
Processing conditions were applied to own voice signals of two talkers selected based on the results presented in Sect.~\ref{sec:pesq_stoi_results}.
One talker represented a low predicted personalization benefit, the other a high predicted benefit.
For each talker, own voice signals were mixed with recorded noise signals (different from the spatialized noise signals mentioned in Sect.~\ref{sec:datasets}, which were used for the evaluation with instrumental metrics).
Four different noise types were considered: surgery noise, metal grinder noise, pseudo-diffuse surgery noise, and pseudo-diffuse factory noise. 
For the surgery noise and metal grinder noise, a noise recording from a real acoustic environment was played back from a loudspeaker 1.5\,m in front of each talker while they were wearing the Hearpiece devices. 
For the pseudo-diffuse surgery noise and the pseudo-diffuse factory noise, time-shifted recordings from real acoustic environments were played back from 8 loudspeakers in 45°-steps with 1.5\,m distance to create an approximately diffuse acoustic environment. 
This recording scenario matches the spatial setup of the spatialized noise signals in Sect.~\ref{sec:datasets}. 
Different from the spatialized setup, these noise signals are recorded and reflect realistic scenarios in which hearables could be used for own voice pickup~\cite{bouserhal2013integration, nordholm_assistive_2015, rennies2023analyse}.
Spectrograms of the recorded noise signals at the outer microphone are shown in Fig.~\ref{fig:specgram_noise}.
The recorded noise signals were also used for evaluation in~\cite{ohlenbusch_multi-microphone_2024} and were only used for testing.

The resulting noisy own voice signals were mixed to an SNR of 0\,dB at the outer microphone. 
This ensured a presentation level of the noisy own voice signal corresponding to 70\,dB sound pressure level (SPL).
For each combination of processing condition, talker, and noise type, utterances of three different sentences were presented to the participants, and the resulting subjective ratings were averaged over utterances.

\section{Results of listening experiment} 
\label{sec:results_listening_exp}

\subsection{Subjective quality ratings}
\label{sec:results_subjective}
\begin{figure*}
    \centering
    \includegraphics[width=\linewidth]{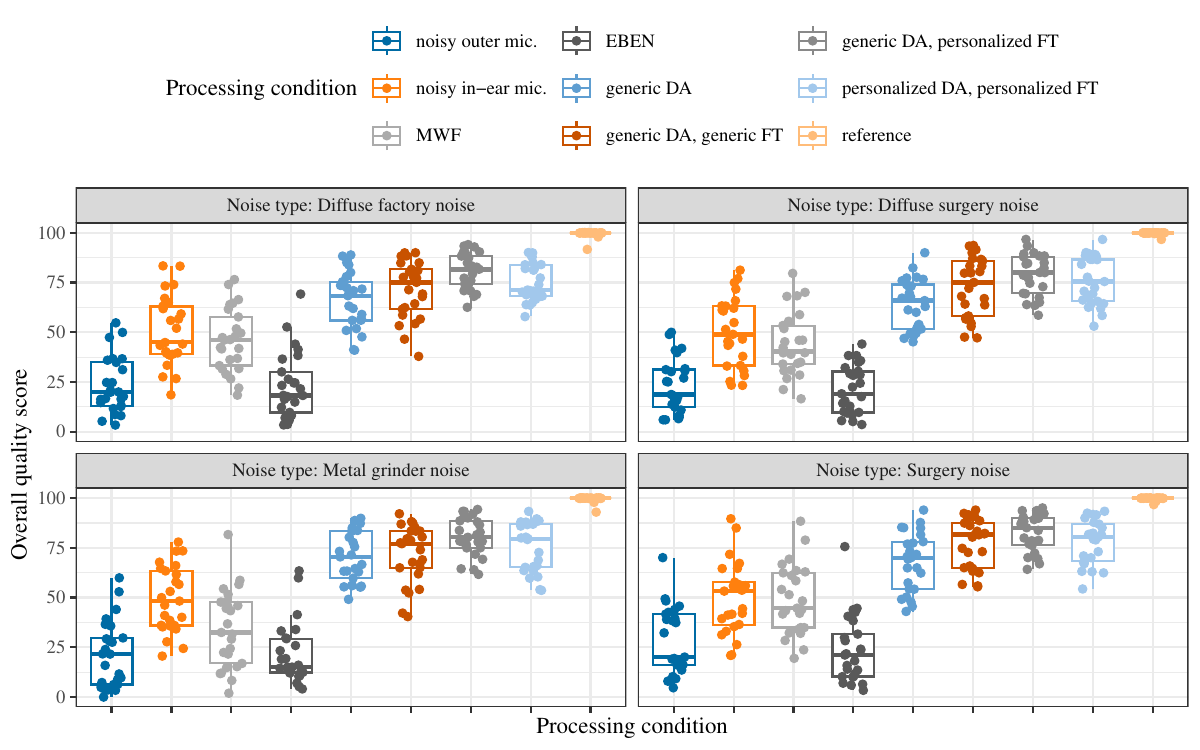}
    \caption{Subjective MUSHRA quality ratings (averaged over sentences) for speech in the \textit{low predicted benefit} case.}
    \label{fig:subj_boxplot_bad_talker}
\end{figure*}

Figure~\ref{fig:subj_boxplot_bad_talker} 
shows the subjective quality ratings for noisy and processed speech in the \textit{low predicted benefit} case. 
Different subplots show the results for speech in different noise types.
The ratings of each condition were very similar across noise types.
In all cases, the clean reference was identified correctly.
The noisy outer microphone signals were rated very low, as expected for the anchor signal, while the noisy in-ear microphone signals were rated close to 50 out of 100 on average. 
MWF and EBEN both performed worse than the noisy in-ear microphone signals.
EBEN in particular was rated similar to the noisy outer microphone signals, indicating no quality improvement. 
All OVR systems trained with the proposed personalization strategies led to considerable improvements over the noisy microphone signals.
In particular, the approaches using personalized fine-tuning achieved the highest quality ratings, closely followed by generic DA, generic FT.

\begin{figure*}
    \centering
    \includegraphics[width=\linewidth]{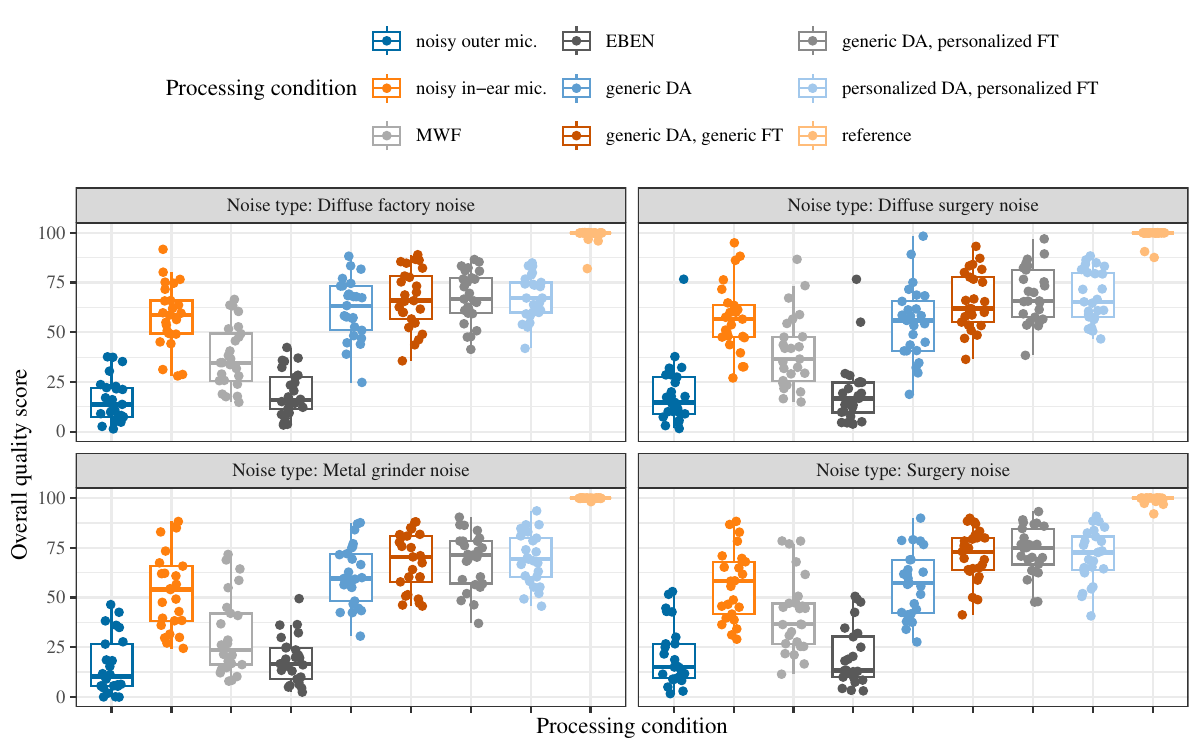}
    \caption{Subjective MUSHRA quality ratings (averaged over sentences) for speech in the \textit{high predicted benefit} case.}
    \label{fig:subj_boxplot_good_talker}
\end{figure*}

Figure~\ref{fig:subj_boxplot_good_talker} shows the subjective quality ratings for noisy and processed speech in the \textit{high predicted benefit} case. 
The results are very similar to those in the \textit{low predicted benefit} case. 
Again, only small differences were observed between systems using generic data augmentation and generic fine-tuning and systems using personalized fine-tuning.
Different from those results, the noisy in-ear microphone was rated slightly better than for speech of the target talker with low predicted benefit. 
In addition, speech processed by the OVR system trained with generic data augmentation without fine-tuning were rated similar to the noisy in-ear microphone signals.


In summary, there is a consistent improvement of speech quality by the proposed personalization strategies that include fine-tuning, compared to both noisy microphone signals and baseline systems. 
The difference observed between the \textit{low predicted benefit} and \textit{high predicted benefit} cases in Sect.~\ref{sec:pesq_stoi_results} was not observed in the subjective ratings.
The OVR systems personalized either by data augmentation or fine-tuning do not yield substantially higher subjective ratings than then generic systems.

\subsection{Statistical analysis of subjective ratings} 
\label{sec:inference_statistics_subj}
Statistical inference was conducted using the R environment~\cite{R_lang}. 
For statistical analysis, subjective ratings were averaged over sentences and noise types.
Shapiro-Wilk tests revealed non-normality of the data.
Visual inspection of QQ-plots indicated saturation effects at both ends of the MUSHRA scale. 
For these reasons, the analysis was carried out by means of non-parametric one-way repeated-measures analyses using separate Friedman rank sum tests for each talker considered. 
Wilcoxon signed rank tests were conducted as post-hoc
tests, and Bonferroni-corrected for multiple comparisons.
The hidden reference ratings were excluded from the statistical analysis.
They served only to anchor participants and to ensure instructions were followed, and thus were not considered a processing condition.

\subsubsection{Low predicted benefit}
\label{sec:stats_lowpred}
In the \textit{low predicted benefit} case, the Friedman test revealed significant differences between processing conditions ($\chi^2(7) = 161.2, p<0.001$).
Post-hoc pairwise comparisons with Bonferroni correction were carried out. The resulting $p$-values are reported in Table~\ref{tab:pvalues_badtalker}. 
The post-hoc tests revealed that the noisy outer microphone signal was rated significantly worse than all other conditions except EBEN, which received similar ratings as the noisy outer microphone signal ($p>0.999$).
The MWF was rated significantly higher than EBEN ($p<0.001$), but not significantly higher from the noisy in-ear microphone ($p>0.999$). 
All generic and personalized approaches based on FT-JNF were rated significantly higher than both noisy microphone signals and significantly higher than both of the baseline systems (MWF and EBEN). 
The system trained with generic data augmentation without fine-tuning was rated significantly worse than the conditions that include fine-tuning. 
Between the conditions that include fine-tuning, the system trained with generic data augmentation and personalized fine-tuning did not perform significantly better than the system trained with generic data augmentation and generic fine-tuning ($p>0.999$).

\begin{table*}
\centering
\caption{Resulting $p$-values from Wilcoxon signed rank test (Bonferroni-corrected) for the \textit{low predicted benefit} case. 
Asterisks indicate significant differences.} 
\label{tab:pvalues_badtalker}
\begingroup
\small
\begin{tabularx}{\textwidth}{Xp{1.4cm}p{1.4cm}p{1.4cm}p{1.4cm}p{1.4cm}p{1.4cm}p{1.4cm}}
  \hline
 & \text{noisy outer} & \text{noisy in-ear} & \text{MWF} & \text{EBEN} & \text{gen. DA} & gen. DA,\newline gen. FT & gen. DA,\newline  pers. FT \\ 
  \hline
noisy in-ear & <0.001* & - & - & - & - & - & - \\ 
  MWF & <0.001* & >0.999 & - & - & - & - & - \\ 
  EBEN & >0.999 & <0.001* & <0.001* & - & - & - & - \\ 
  gen. DA & <0.001* & ~\,~0.002* & <0.001* & <0.001* & - & - & - \\ 
  gen. DA, gen. FT & <0.001* & <0.001* & <0.001* & <0.001* & ~\,~0.001* & - & - \\ 
  gen. DA, pers. FT & <0.001* & <0.001* & <0.001* & <0.001* & <0.001* & <0.001* & - \\ 
  pers. DA, pers. FT & <0.001* & <0.001* & <0.001* & <0.001* & <0.001* & >0.999 & <0.001* \\ 
   \hline
\end{tabularx}
\endgroup
\end{table*}

\subsubsection{High predicted benefit}
\label{sec:stats_highpred}
In the \textit{high predicted benefit} case, the Friedman test also revealed significant differences between processing conditions ($\chi^2(7)=151.34, p<0.001$). 
Posthoc pairwise comparisons with Bonferroni correction were carried out. The resulting $p$-values are reported in Table~\ref{tab:pvalues_goodtalker}.
Similar to the results in the \textit{low predicted benefit} case, here the noisy outer microphone signal was rated significantly worse than all other processing conditions except EBEN ($p>0.999$).
The MWF was rated significantly higher than EBEN ($p<0.001$), but not significantly higher from the noisy in-ear microphone signal ($p>0.999$).
All generic and personalized approaches based on FT-JNF were rated significantly higher than both noisy microphone signals and significantly higher than both of the baseline systems (MWF and EBEN), except for the system trained with generic data augmentation and without fine-tuning, which was not rated significantly different from the noisy in-ear microphone signals ($p>0.999$).
The systems trained with fine-tuning were rated significantly higher than the system trained with generic data augmentation and without fine-tuning. 
Between the fine-tuned systems, there were no significant differences.

\begin{table*}
\centering
\caption{Resulting $p$-values from Wilcoxon signed rank test (Bonferroni-corrected) for the \textit{high predicted benefit} case. 
Asterisks indicate significant differences.} 
\label{tab:pvalues_goodtalker}
\begingroup\small
\begin{tabularx}{\textwidth}{Xp{1.4cm}p{1.4cm}p{1.4cm}p{1.4cm}p{1.4cm}p{1.4cm}p{1.4cm}}
  \hline
 & \text{noisy outer} & \text{noisy in-ear} & \text{MWF} & \text{EBEN} & \text{gen. DA} & gen. DA,\newline gen. FT & gen. DA,\newline  pers. FT \\ 
  \hline
noisy in-ear & <0.001* & - & - & - & - & - & - \\ 
  MWF & <0.001* & <0.001* & - & - & - & - & - \\ 
  EBEN & >0.999 & <0.001* & <0.001* & - & - & - & - \\ 
  gen. DA & <0.001* & >0.999 & <0.001* & <0.001* & - & - & - \\ 
  gen. DA, gen. FT & <0.001* & ~\,~0.001* & <0.001* & <0.001* & <0.001* & - & - \\ 
  gen. DA, pers. FT & <0.001* & <0.001* & <0.001* & <0.001* & <0.001* & >0.999 & - \\ 
  pers. DA, pers. FT & <0.001* & <0.001* & <0.001* & <0.001* & ~\,~0.005* & >0.999 & >0.999 \\ 
   \hline
\end{tabularx}
\endgroup
\end{table*}

\subsection{Prediction of subjective quality ratings} 
\label{sec:pred_subj_ratings}
This section investigates how well the objective metrics described in Sect.~\ref{sec:objective_metrics} predict the subjective ratings described in Sect.~\ref{sec:results_subjective}. The prediction performance was measured in terms of correlation and root mean squared error (RMSE) between objective metric prediction and the median subjective quality rating.
In particular, the Pearson linear correlation coefficient $r$ was used to assess the accuracy of the predictions, and the Spearman rank coefficient $\rho_S$ was used to assess the monotonicity of the predictions. 
The same audio signals as in the subjective evaluation were used to compute predictions for each individual signal. For each of the eight processing conditions, 24 predictions (two cases, three sentences, four noise types) were compared with the corresponding subjective ratings (median over subjects).
We computed RMSE values for raw predictions(see e.g.,~\cite{beerends2019subjective}). 
In addition, RMSE was computed after fitting a thrid-order polynomial (RMSE$_\text{3}$) to account for systematic variation in subjective ratings
For computing RMSE values, the predictions were scaled to the MUSHRA range to facilitate comparisons between metrics\footnote{It should be noted that some of the DNN-based metrics produced values outside their respective range, e.g., WV-MOS predictions are supposed to lie on the MOS scale between 1 and 5, but negative predictions were observed.}.
For PEMO-Q, an output range between 0 and 1 was considered.
For LEAP, an output range between 1 and 13 was considered.
For SCOREQ distance, no RMSE value is reported, since the distance scale is open-ended. 
This scaling procedure assumes a linear relationship between the (relative) MUSHRA scale and the (absolute) scales of the metrics, such as the MOS scale.

\begin{figure*}
    \centering
    \includegraphics[width=0.9\textwidth]{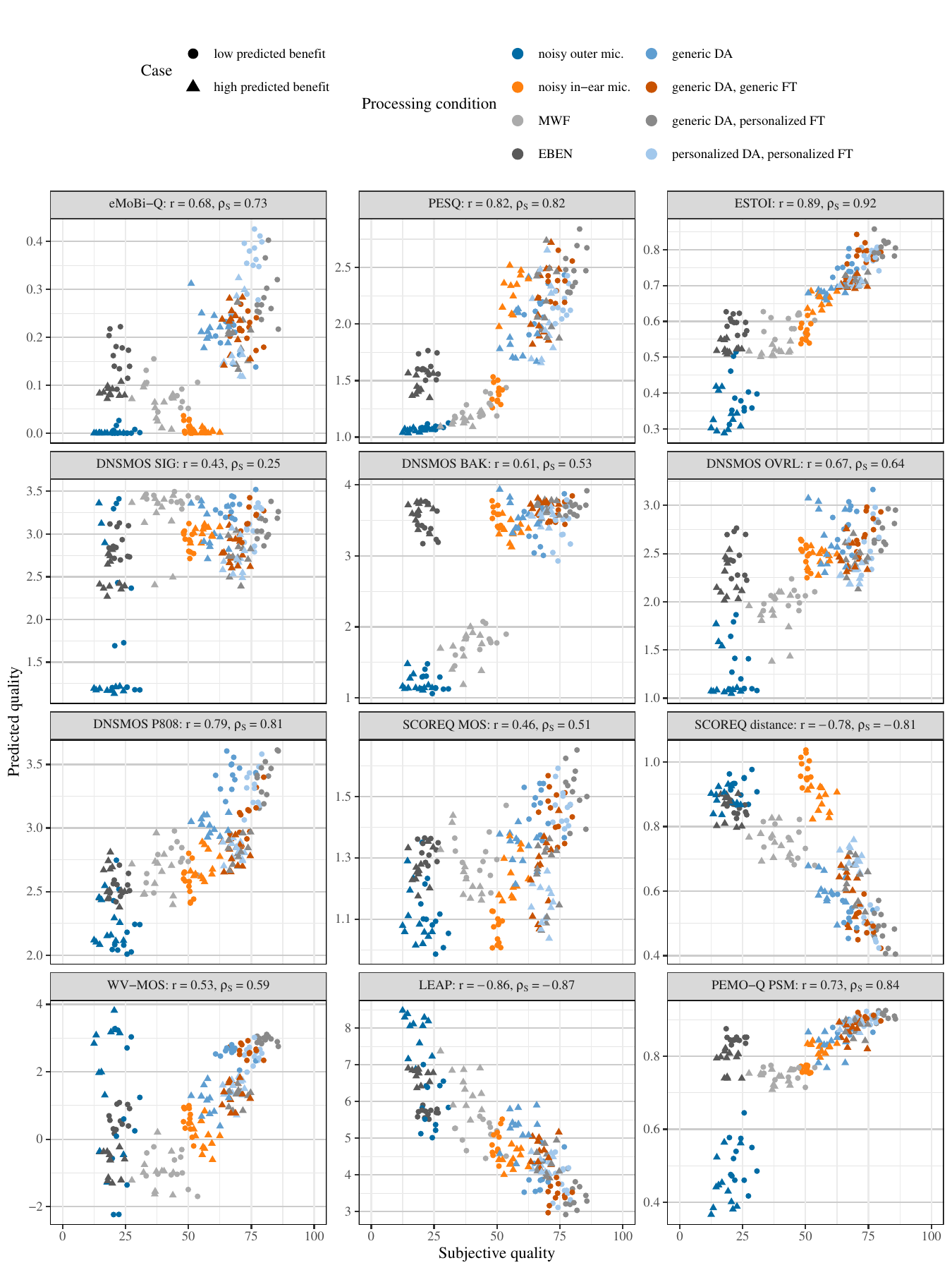} 
    \caption{Scatterplots comparing median subjective quality ratings (abscissae) and predicted objective quality (ordinates) for noisy and reconstructed own voice signals.}
    \label{fig:correlation_plot}
\end{figure*}
The prediction performance of the objective metrics is shown in Fig.~\ref{fig:correlation_plot}.
Different symbols correspond to the own voice signals of different talkers, while different colors correspond to different processing conditions. 
For each metric, the quality predictions of different processing conditions can be observed to group in clusters.
Most metrics predicted higher quality for conditions with higher subjective ratings. 
In the case of DNSMOS~SIG and DNSMOS~BAK, there is a high spread of ratings that does not seem to follow a strong common trend. 
In the case of SCOREQ~distance, there appears to be a strong, but negative correlation, since for the distance-based metric lower values indicate higher similarity. 
The SCOREQ~distance values for the noisy in-ear microphone signals are similar to the noisy outer microphone signals or the signals processed by EBEN, despite the noisy in-ear microphone signals receiving much higher subjective ratings. 

LEAP achieved a strong negative correlation, indicating that lower listening effort corresponds to higher subjective quality.  
The majority of the metrics consistently overestimated the quality of the signals processed by EBEN with respect to the other processing conditions (see clusters of black symbols in each panel).  
In contrast, the quality of the noisy in-ear signals was consistently underestimated by eMoBi-Q and the SCOREQ~MOS and distance predictions. 
The bandwidth extension-based metric WV-MOS strongly overestimated the quality of the full-bandwidth noisy outer microphone signals.
Comparing the predictions for low and high predicted benefit, the predictions by all metrics tend to cluster together for both cases, with the individual processing conditions forming distinct clusters. 
This indicates that the processing condition has a larger influence on the predictions than the case (low or high predicted benefit), which was consistently observed in the experiments as well as in most predictions.  

\begin{table}
\centering
\caption{\label{tab:correlations_and_rmse}Correlation coefficients and RMSE values for the considered metrics.}
\centering
\begin{tabular}[t]{lrrrr}
\toprule
\textbf{Metric} & \textbf{$r$} & \textbf{$\rho_S$} & \textbf{RMSE} & \textbf{RMSE$_\text{3}$}\\
\midrule
eMoBi-Q & 0.68 & 0.73 & 40.8 & 15.2\\
PESQ & 0.82 & 0.82 & 23.5 & 11.8\\
ESTOI & 0.89 & 0.92 & 16.0 & 8.7\\
DNSMOS SIG & 0.43 & 0.25 & 20.7 & 18.4\\
DNSMOS BAK & 0.61 & 0.53 & 19.8 & 16.8\\
DNSMOS OVRL & 0.67 & 0.64 & 25.2 & 15.6\\
DNSMOS P808 & 0.79 & 0.81 & 16.7 & 12.5\\
SCOREQ MOS & 0.46 & 0.51 & 49.2 & 18.6\\
SCOREQ distance & -0.78 & -0.81 & - & 11.9\\
WV-MOS & 0.53 & 0.59 & 58.6 & 16.7\\
LEAP & -0.86 & -0.87 & 36.2 & 10.3\\
PEMO-Q PSM & 0.73 & 0.84 & 31.3 & 12.6\\
\bottomrule
\end{tabular}
\end{table}
Correlation coefficients as well as root mean squared error (RMSE) values are reported in Table~\ref{tab:correlations_and_rmse}.
In terms of linear correlation, ESTOI predictions achieved the highest (absolute) correlation coefficient, followed by LEAP, 
PESQ, PEMO-Q~PSM, DNSMOS~P808, and SCOREQ distance. 
The absolute correlation achieved by SCOREQ distance is almost as high as for P808.
Slightly lower correlations were achieved by eMoBi-Q and DNSMOS~OVRL.
DNSMOS~BAK, WV-MOS, SCOREQ~MOS and DNSMOS~SIG achieved the lowest correlations out of the metrics considered. 
In terms of rank correlation, the order of metrics is similar, with the correlation of ESTOI predictions being particularly high ($\rho_S=0.92$). 
For most metrics, the linear and rank correlations are very close, except for DNSMOS~SIG where a much lower rank correlation ($\rho_S=0.25$) was achieved than linear correlation ($r=0.43$).
In terms of RMSE, the residual errors of the predicted values are generally higher compared to the RMSE values reported in, e.g.,~\cite{eurich_computationally_2024}.
As visible in Fig.~\ref{fig:correlation_plot}, many of the metrics do not predict close to the upper end of their respective scale for processing conditions that were rated to have very high subjective quality or do not reach the lower end of their respective scale for processing conditions rated to have very low subjective quality. 
The lowest RMSE was achieved by ESTOI, while the highest RMSE was achieved by WV-MOS.
The RMSE$_\text{3}$ values after fitting a third-order polynomial are generally lower than the RMSE values computed from the raw predictions. In particular, the values for ESTOI are the lowest, followed by LEAP, PESQ, and SCOREQ~distance.

\section{Discussion}
\label{sec:discussion}
\subsection{Comparison to previous research}
\label{sec:disc_comparison_sota}
Previous research on generic OVR systems has compared different approaches using subjective ratings, e.g.,~\cite{hauret_configurable_2023, li_two-stage_2024}.
In~\cite{li_two-stage_2024}, objective metrics (PESQ and STOI) and subjective ratings showed an improvement through OVR processing.
While in~\cite{li_two-stage_2024} objective metrics and subjective ratings were not compared in terms of correlation, a comparison of objective metrics and subjective ratings in~\cite{hauret_configurable_2023} revealed low correlation of PESQ and SI-SDR (a technical metric) with subjective ratings, but higher correlations were observed for STOI and Noresqa-MOS (a reference-free metric).
However, the experiments in~\cite{hauret_configurable_2023} were carried out using only simulated own voice signals and only generic systems were evaluated. 
In comparison, the results presented in this paper show a higher correlation for PESQ with subjective quality ratings than in~\cite{hauret_configurable_2023}, while ESTOI achieved similarly high correlations as STOI in~\cite{hauret_configurable_2023}.

In both~\cite{li_two-stage_2024} and~\cite{hauret_configurable_2023}, OVR was only performed using single-channel body-conduction sensor signals as input, whereas the multi-channel OVR systems in this work use both an in-ear and an outer microphone as input. 
Additionally, personalized OVR systems are considered, and recorded own voice signals are used for evaluation.

Although previous research has already investigated personalization of OVR systems, there has not been any subjective evaluation of personalized OVR systems as far as we know. 
In~\cite{edraki_speaker_2024}, personalized OVR systems using an in-ear microphone were proposed, but only evaluated in terms of objective metrics. 
Due to the proposed personalization method, it was observed that the generalization to talkers not in the training data was poor. 
Similarly, in~\cite{sui_tramba_2024} personalized OVR systems were proposed, but the benefit of personalization was not assessed in terms of subjective ratings.
In~\cite{he_towards_2023}, personalization was achieved by calibrating an OVR system with a few minutes of recorded speech signals from the target talker, but the benefit was not assessed in terms of subjective ratings either.
This paper addresses these knowledge gaps by comparing personalized OVR systems using both subjective and objective evaluations.

In terms of personalized data augmentation, previous work has already investigated data augmentation based on text-to-speech systems for synthesizing training data for training personalized speech enhancement systems, e.g., in~\cite{Kuznetsova_2023_speechsynth_pse, bae2025generative}.
These approaches also perform personalized data augmentation, the augmentation is based on synthesizing speech signals for the speech production characteristics of a specific talker, such as prosody and pitch.
In~\cite{Kuznetsova_2023_speechsynth_pse}, speech synthesis-based data augmentation was able to improve personalized speech enhancement performance compared to a generic system.
In~\cite{bae2025generative}, zero-shot text-to-speech systems were used to augment data for training personalized speech enhancement systems, which outperformed generic systems.
Differently, this paper investigates the significance of simulating transfer characteristics of own voice between hearable microphones for personalized data augmentation in order to train multi-channel OVR systems, addressing a gap in previous research.

%
While a direct comparison of text-to-speech-based data augmentation with the methods is outside the scope of this paper, it could be interesting to compare or even combine these techniques in future work.
As a possible topic for further research, it would be interesting to investigate training personalized speech enhancement systems conditioned on auxiliary information about the target talker (see e.g.,~\cite{ZmolikovaOverview2023}) on a larger dataset with a sufficient amount of different talkers, such as Vibravox~\cite{jhauret-et-al-2025-vibravox}, and to compare between different approaches to personalization.

\subsection{Comparison of OVR systems}
\label{sec:disc_comparison_ovr}
The subjective listening test showed that OVR systems improved quality ratings over both noisy in-ear and noisy outer microphone signals.
In addition, the OVR systems yield better performance than the considered baselines EBEN and MWF.

%
While personalized OVR was predicted to have a consistent benefit by instrumental metrics, this benefit was not consistent over all conditions in the results of the subjective listening test.
The best performance is obtained when OVR systems are trained with generic data augmentation and personalized fine-tuning, although in some conditions there is no gain from personalized over generic fine-tuning, indicating that the benefits of personalization may be situation-dependent or limited.

\subsection{OVR quality prediction}
\label{sec:disc_quality_pred}
In terms of quality prediction by instrumental metrics, the results have shown that not all metrics were able to predict subjective quality for OVR systems.
Among the reference-based metrics, ESTOI, PESQ, and PEMO-Q~PSM achieved the highest correlations. 
Among the reference-free metrics, LEAP, SCOREQ~distance, and DNSMOS~P808 achieved the highest correlations.
In most cases, most metrics deviated from subjective ratings when predicting the quality of noisy or bandwidth-extended in-ear signals.
This observation matches previously published research on subjective evaluation of bandwidth-extended~\cite{Bauer2015artificial} or body-conducted speech~\cite{richard_comparison_2023}. 
In~\cite{pulakka15_interspeech} it was observed that both PESQ and POLQA were unable to correctly predict ranking of bandwidth extension algorithms. 
This was also observed in Fig.~\ref{fig:correlation_plot} for the predictions of PESQ for the signals processed by EBEN.
For in-ear microphone signals, in~\cite{santos2016objective} the quality predictions obtained with PESQ were highly correlated with subjective ratings, which is consistent with the observations in this paper.

Other DNN-based metrics also exhibit deviations from subjective ratings of noisy or bandwidth-extended signals, which could be attributed to the fact that in-ear microphone signals are subject to degradations like individually different bandwidth limitations, body-produced noise, or time-varying changes in transfer characteristics. 
These degradations are not typical in standard single-channel speech enhancement evaluations as considered for the training of e.g., DNSMOS~\cite{reddy_dnsmos_2022}. 
While WV-MOS was reported to achieve much higher correlation than PESQ or DNSMOS metrics for prediction of MOS for bandwidth extension in~\cite{andreev_hifipp_2023_arxiv}\footnote{Note that this particular result is only reported in the extended preprint~\cite{andreev_hifipp_2023_arxiv} and not in the conference paper~\cite{andreev_hifipp_2023}.}, the results in this paper do not reflect this. 
A possible explanation is the difference between signals band-limited by a lowpass filter and own voice signals recorded by an in-ear microphone.
The high correlations of LEAP are somewhat surprising because the model was developed for listening effort predictions and not for quality predictions. 
The ASR system underlying the phoneme classification had never seen strong band limitation or bandwidth extension algorithms during training. 
This may indicate the phoneme classification-based perception prediction generalizes well across different signal degradations and enhancement strategies, as suggested in~\cite{Rennies2025} for different types of signal degradations.

\section{Conclusion}
\label{sec:conclusion}
In this paper, we have investigated personalized own voice reconstruction systems and evaluated their performance in terms of subjective quality. 
The systems were personalized during training with augmented data, or during fine-tuning with recorded data. 
Personalized systems were compared to their generic counterparts.
While objective metrics predicted an increase in quality through personalization, the subjective evaluation only partly confirmed this improvement.
In particular, most metrics failed to accurately predict the quality of band-limited noisy or bandwidth-extended in-ear microphone signals.
This mismatch was observed when comparing objective predictions with median subjective ratings.
Other metrics, such as the intrusive ESTOI and the non-intrusive LEAP, correlated well with subjective ratings. 
Both objective and subjective results demonstrate that the considered own voice reconstruction systems substantially increase own voice quality compared to noisy microphone signals and several baseline systems.

\section*{Author declarations}
\label{sec:author_declarations}

\conflict 
\noindent The authors declare no conflict of interest.

\informedconsent
\noindent All subjects who participated in the recordings were informed about data collection and future data use, and gave informed consent.

\dataavailability
\noindent The research data associated with this article are available in Zenodo, under the references \url{https://doi.org/10.5281/zenodo.10844598} (recorded own voice signals)~\cite{ohlenbusch_2024_zenodo_German}, \url{https://doi.org/10.5281/zenodo.11196866} (transfer function measurements)~\cite{ohlenbusch_2024_zenodo_Transfer},
and \url{https://doi.org/10.5281/zenodo.15248719} (subjective ratings, objective predictions, and audio signals of the listening experiment stimuli)~\cite{ohlenbusch_2025_zenodo}.
Selected audio examples are also available at \url{https://m-ohlenbusch.github.io/subjective_ovr_personalized/}.

\acknowtext
\noindent The Oldenburg Branch for Hearing, Speech and Audio Technology HSA is funded in the program \frqq Vorab\flqq~by the Lower Saxony Ministry of Science and Culture (MWK) and the Volkswagen Foundation for its further development.
This work was partly funded by the German Ministry of Science and Education BMBF FK 16SV8811 and the Deutsche Forschungsgemeinschaft (DFG, German Research Foundation) - Project ID 352015383 - SFB 1330 C1 and A1.
We thank Fenja Hermann for conducting the listening experiment and Rainer Huber for helpful discussion and computing the LEAP and PEMO-Q~PSM predictions.
We also thank the study subjects for their participation in the listening experiment.

\printbibliography 

\end{document}